\documentclass[aps,prl,preprint, showpacs]{revtex4}

\usepackage{epsfig}
\usepackage{graphicx}

\def\be{\begin{equation}}
\def\ee{\end{equation}}
\def\bea{\begin{eqnarray}}
\def\eea{\end{eqnarray}}

\begin{document}
\draft

\title{Electron spin operation by electric fields:\\
spin dynamics and spin injection}
\author{Emmanuel I. Rashba\cite{Rashba*}}
\affiliation{Department of Physics, SUNY at Buffalo, Buffalo, New York 14260, USA}
\date{\today,}

\begin{abstract}
Spin-orbit interaction couples electron spins to electric fields and allows electrical monitoring of electron spins and electrical detection of spin dynamics. Competing mechanisms of spin-orbit interaction are compared, and optimal conditions for the electric operation of electrons spins in a quantum well by a gate voltage are established. Electric spin injection into semiconductors is discussed with a special emphasis on the injection into ballistic microstructures. Dramatic effect of a long range Coulomb interaction on transport phenomena in space-quantized low-dimensional conductors is discussed in conclusion.

\end{abstract}

\maketitle

\label{intro}

Manipulating electron spins at a given location and transporting electron spins between different locations belong to the central problems of the growing field of semiconductor spintronics \cite{Wolf,DS01} that is of critical importance for quantum computing and information processing.

Most of computing schemes with electron spins in semiconductor microstructures are based on using time dependent magnetic fields. However, from experimental point of view, using time-dependent electric fields instead of magnetic ones is highly preferable. It would allow local access, at a few nanometer scale, to individual microstructures. Existence of various mechanisms of spin-orbit (SO) coupling \cite{RS91} opens attractive possibilities for electrical control of electron spins, especially because electron confinement in low-symmetry environment enhances SO coupling and produces new mechanisms of it. Potentialities of electrical spin manipulation are recognized already, and successful experiments have been reported recently \cite{Kato}. We identify SO coupling mechanisms that are most promising for the gate voltage control of spin dynamics in narrow gap semiconductors and show that the dependence of the electric dipole spin resonance (EDSR) on the magnetic field direction is a unique method for identifying different SO coupling mechanisms contributing to EDSR in quantum wells \cite{RE}. Electric signals produced by precessing electron spins can be used for noninvasive single spin detection by electrical techniques \cite{LR03}. 

Transport of electron spins between different locations, either between quantum dots \cite{LDV} or from a ferromagnetic spin injector into a nonmagnetic conductor \cite{AP76,JS85}, is also monitored by electric fields. Efficiency of spin injection depends on the conductivities of both conductors and the properties of a contact between them. Because a spin transistor \cite{DD90} and similar devices rely mostly on the ballistic transport inside a semiconductor microstructure \cite{Loss93} while transport in metallic leads is always diffusive, a criterion for efficient spin injection across a diffusive-ballistic-diffusive junction is of special interest. We show that it is the Sharvin resistance \cite{Sh65} of the ballistic region that controls spin injection into it \cite{KR03}. Fast switching time required of  nanoelectronic devices makes the reactive part of their impedance no less important than the active part of it. In low-dimensional spatially-quantized conductors electrical screening is suppressed. As a result, electrical capacitance of inhomogeneities can grow strongly and transients should be influenced by these anomalies \cite{KR02}.

\section{Spin-orbit coupling in semiconductor microstructures}
\label{SO}

Free and weakly bound electrons in semiconductors experience SO coupling that is much stronger than in a vacuum. Because this enhancement in the SO coupling strength is of critical importance for the electric operation of electron spins and its applications in spintronics, we outline here briefly the enhancement mechanism. For a slow electron in a vacuum, the Hamiltonian of SO interaction reads
\begin{equation}
 H_{\rm so}(\mbox{\boldmath $r$})={{e\hbar}\over{4m_0c^2}}\mbox{\boldmath$\sigma$}\cdot(\mbox{\boldmath$E$}(\mbox{\boldmath$r$})\times\mbox{\boldmath$v$}),
\label{eq1a}
\end{equation}
 where \mbox{\boldmath$\sigma$} is the Pauli matrix vector and \mbox{\boldmath$E$}(\mbox{\boldmath$r$}) is an electric field. For a slow electron, $v/c\ll1$, and a weak field \mbox{\boldmath$E$}(\mbox{\boldmath$r$}), $H_{\rm so}(\mbox{\boldmath$r$})$ is small because of the Dirac gap in the denominator, $2m_0c^2\approx $ 1 MeV. Presence of the field \mbox{\boldmath$E$}(\mbox{\boldmath$r$}) in Eq.~(\ref{eq1a}) suggests that an electron possesses a dipole moment
\begin{equation}
 \mbox{\boldmath$P$}(\mbox{\boldmath$v$})={{e\hbar}\over{4m_0c^2}}(\mbox{\boldmath$\sigma$}\times\mbox{\boldmath$v$}),
\label{eq2a}
\end{equation}
that is small for the reasons explained above and vanishes when $v\rightarrow 0$.

There are fundamental symmetry requirements behind the fact that the magnetic moment \mbox{\boldmath$\mu$} of an electron is related to its spin $\mbox{{\boldmath$s$}}=\hbar\mbox{\boldmath$\sigma$}/2$ by a simple relation
\begin{equation}
 \mbox{\boldmath$\mu$}=(e/m_0c)\mbox{\boldmath$s$}
\label{eq3a}
\end{equation}
where the coefficient $e/m_0c$ is a gyromagnetic ratio, while the electric dipole moment \mbox{\boldmath$P$} involves electron velocity \mbox{\boldmath$v$}. These requirements are based on the space and time inversion symmetries. With respect to the space inversion, \mbox{\boldmath$s$} and \mbox{\boldmath$\mu$} behave as pseudovectors while \mbox{\boldmath$v$} and \mbox{\boldmath$P$} as vectors. With respect to the time inversion, \mbox{\boldmath$s$}, \mbox{\boldmath$\mu$}, and \mbox{\boldmath$v$} change sign and are considered as imaginary quantities, while \mbox{\boldmath$P$} remains unchanged and is considered as a real quantity. Because \mbox{\boldmath$E$} is a real vector and a magnetic field \mbox{\boldmath$B$} is an imaginary pseudovector, one can easily check that Eqs.~(\ref{eq1a}) - (\ref{eq3a}) are invariant with respect to both space and time inversions. Strong Zeeman coupling $H_Z=-(\mbox{\boldmath$\mu$}\cdot\mbox{\boldmath$B$})$ of the electron spin to $\mbox{\boldmath$B$}$ and weak coupling to $\mbox{\boldmath$E$}$ directly follow from the above symmetry arguments.

Enhancement of SO coupling in solids comes from two basic sources. First, this coupling originates mostly from the fast electron motion in a strong electric field near nuclei rather than from the translational motion with a small velocity \mbox{\boldmath$v$}. The effect of this coupling is built-in in periodic parts $u_{\mbox{\boldmath$k$}}(\mbox{\boldmath$r$})$ of the Bloch functions $\psi_{\mbox{\boldmath$k$}}(\mbox{\boldmath$r$})=u_{\mbox{\boldmath$k$}}(\mbox{\boldmath$r$})
\exp[i(\mbox{\boldmath$k$}\cdot\mbox{\boldmath$r$})]$, and it manifests itself in a large SO splitting of the valence bands, $\Delta\sim 1$ eV, that is typically comparable to or even larger than a forbidden gap $E_G$. Mathematically, in the spirit of the Kane scheme, equations of the band theory of narrow gap semiconductors are similar to a Dirac equation but with the forbidden gap $E_G$ instead of the Dirac gap $2m_0c^2$. This difference in equations reflects enormous increase in SO interaction in narrow gap semiconductors as compared to a vacuum. SO effects in narrow-gap and zero-gap quasi-2D systems were investigated analytically, numerically, and experimentally \cite{Rad96,Beck01,ZP}.

Second, the symmetry of crystals, and especially the symmetry of microstructures, is essentially lower than the symmetry of a vacuum. As a result, new terms that critically change spin dynamics appear in electron Hamiltonians. In systems lacking an inversion center, the two-fold degeneracy of quantum states with a given momentum \mbox{\boldmath $k$} ($\vert{\mbox{\boldmath $k$}}\uparrow\rangle$ and $\vert{\mbox{\boldmath $k$}}\downarrow\rangle$) is not required throughout the Brillouin zone. In fact, it is lifted everywhere with the exception of the point \mbox{{\boldmath $k$}=0} and some symmetry lines (or symmetry planes). Such changes in the Hamiltonians and the energy spectra enhance the effect of the SO coupling. E.g., in uniaxial systems like hexagonal crystals of the CdS type and asymmetric quantum wells, linear in \mbox{\boldmath$k$} terms develop in electron Hamiltonians. These terms result in a linear in \mbox{\boldmath$k$} zero-magnetic-field splitting of the spectrum, and the electric dipole moment for spin-flip transitions equals
\begin{equation}
 \mbox{\boldmath$P$}(\omega_s)=i{{e\alpha}\over{\hbar\omega_s}}(\mbox{\boldmath$\sigma$}\times\hat{\mbox{\boldmath$z$}}).
\label{eq4a}
\end{equation}
Here $\hat{\mbox{\boldmath$z$}}$ is a unit vector in the direction of the symmetry axis, and $\alpha$ is a SO coupling constant, cf. Eq.~(\ref{eq3}) below. This equation can be derived by substituting in the electron Hamiltonian the canonical momentum \mbox{\boldmath$k$} by a kinetic momentum $\hat{\mbox{\boldmath $k$}}=-i\mbox{\boldmath$\nabla$}-e\mbox{\boldmath$A$}(\mbox{\boldmath$r$},t)/\hbar c$, where $\mbox{\boldmath$A$}(\mbox{\boldmath$r$},t)$ is a vector potential. Expressing the time dependent part of \mbox{\boldmath$A$} through the electric field \mbox{\boldmath$E$}(t) we come to Eq.~(\ref{eq4a}). The imaginary unit $i$ in this equation ensures the proper symmetry with respect to the time inversion. Because of the spin-flip energy $\hbar\omega_s(B)$ in the denominator of Eq.~(\ref{eq4a}), the electric dipole $\mbox{\boldmath$P$}(\omega_s)$ is usually large. In the $B\rightarrow 0$ limit the denominator is equal to the zero-field splitting, $\hbar\omega_s=2\alpha k$, and the SO coupling constant $\alpha$ cancels from the right hand side of Eq.~(\ref{eq4a}). Hence, under these conditions the electric dipole is especially large. We emphasize that in the absence of an external magnetic field,  $\mbox{\boldmath$B$}=0$, the time inversion symmetry is preserved, hence, an electron possesses no static electric dipole moment. It is the dipole moment of the transition at the zero-field splitting frequency that does not vanish and is large.

Therefore, low-symmetry narrow-gap systems formed from heavy chemical elements are best candidates for a strong SO coupling. In what follows, we concentrate on the coupling of electron spins to electric fields, static and dynamic. In particular, SO coupling allows to perform Rabi oscillations between up- and down-spin states using time dependent electric, rather than magnetic, fields \cite{R60}.

Several examples of SO coupling follow. For electrons in direct-gap A$_3$B$_5$ compounds, the SO contribution to a total Hamiltonian $H$ is \cite{D55,RS61}

\begin{equation}
 H_{3D}=\delta(\mbox{\boldmath $\sigma$}\cdot\hat{\mbox{\boldmath $\kappa$}}),
\label{eq1}
\end{equation}
where ${\hat\kappa}_x={\hat k}_y{\hat k}_x{\hat k}_y-{\hat k}_z{\hat k}_x{\hat k}_z$, ${\hat \kappa}_y$ and ${\hat \kappa}_z$ can be derived from ${\hat \kappa}_x$ by cyclic permutations, and $\hat{\mbox{\boldmath $k$}}$ is the kinetic momentum. This Hamiltonian is known as the Dresselhaus Hamiltonian. The constant $\delta$ is about 20 eV \AA$^3$ for GaAs and about 150 to 250 eV \AA$^3$ for InAs, InSb, and GaSb. For a narrow [0,0,1] quantum well, the Hamiltonian of Eq.~(\ref{eq1}) reduces to the 2D Dresselhaus term \cite{LMR85,BR85,DK86}
\begin{equation}
 H_D=\alpha_D(\sigma_x{\hat k}_x-\sigma_y{\hat k}_y),~~\alpha_D=-\delta(\pi/w)^2,
\label{eq2}
\end{equation}
$w$ being the quantum well width. With $w=100$ \AA, $\alpha_D$ ranges from about $2\times 10^{-10}$ eV cm to $2\times 10^{-9}$ eV cm and decreases rapidly with $w$.

While the $H_D$ term comes from the bulk inversion asymmetry (BIA), there exists another contribution
\begin{equation}
 H_R=\alpha_R(\sigma_x{\hat k}_y-\sigma_y{\hat k}_x)
\label{eq3}
\end{equation}
known as the Rashba term. In wurtzite type crystals it develops because of BIA \cite{RS91}, but in A$_3$B$_5$ quantum wells it appears due to a structure inversion asymmetry (SIA) \cite{BR84} that can be changed by applying an electric field $E$ across the quantum well. For InAs based quantum wells, typical values of $\alpha_R$ are about $\alpha_R\sim10^{-9}$ eV cm, however, values as large as $\alpha_R\approx6\times10^{-9}$ eV cm have been also reported \cite{Cui}. There is no simple way to calculate $\alpha_R$ because it depends both on the field $E$ inside the quantum well and the boundary conditions at the interfaces \cite{PZ98,Maj}. The importance of this interaction stems from the fact that $\alpha_R$ can be controllably changed by a gate voltage as it was shown as applied to InAs based quantum wells by Nitta {\it et al.} \cite{Nitta} and Engels {\it et al.} \cite{Eng97}, and also in a number of more recent papers. This property makes SIA a prospective candidate for developing a spin transistor \cite{DD90}.

One more orbital mechanism of SO interaction stems from the ``anomalous" SO contribution to the coordinate operator $\hat{\mbox{\boldmath $r$}}_{\rm so}={\it l}^2_{\rm so}(\mbox{\boldmath $\sigma $}\times\hat{\mbox{\boldmath $k$}})$ \cite{Yafet} rather than from the SO part of the Hamiltonian. For narrow gap semiconductors the length ${\it l}_{\rm so}$ can be estimated in the framework of the Kane model as ${\it l}_{\rm so}\approx\hbar(|g|/4m_0E_G)^{1/2}$, where $g$ is a $g$-factor. McCombe {\it et al.} have shown that this mechanism is rather efficient in bulk n-InSb \cite{MBK}, however, only at the combinational, spin-cyclotron, frequency. In asymmetric quantum wells this mechanism should produce transitions also at the spin flip frequency $\omega_s$. Measuring the magnitude of this operator in a realistic environment is an important and challenging task.

While the three above mechanisms of SO coupling stem from the electron orbital motion, the mechanism employed by Kato {\it et al.} \cite{Kato} originates from the spatial dependence of the $\hat g$ tensor, ${\hat g}={\hat g}(z)$. Under these conditions a Zeeman Hamiltonian $H_Z=\mu_B(\mbox{\boldmath $\sigma $}\hat{g}(z)\mbox{\boldmath $B$})/2$ involves both the coordinate $z$ and Pauli matrices, hence, it includes SO orbit coupling in itself. This mechanism is most efficient when the $\hat g$-tensor is  small, $|g|\sim 0.1$, and highly anisotropic. Wide parabolic AlGaAs quantum wells used in Ref.~\cite{Kato} are optimal for this mechanism. In the next section, we concentrate on the SO coupling mechanisms that are efficient for large $g$-factors typical of narrow gap semiconductors.

\section{Gate voltage manipulation of electron spins}
\label{gate}

A gate voltage $V(t)$ applied to a quantum well produces a perpendicular-to-plane electric field ${\tilde E}(t)$ and a time dependent perturbation $H(t)=e{\tilde E}(t)z$. In a framework of strictly 2D models the operator $z$ commutes with $H$, hence, it cannot influence spin dynamics. Therefore, a gate voltage control of spin dynamics is possible only due to a deviation of the quantum well from the strictly 2D limit. A parabolic quantum well with a confinement potential $H_{\rm conf}=m\omega_0^2z^2/2$ is the only model that can be solved analytically and allows considering a tilted magnetic field that mixes the in-plane and perpendicular-to-plane motions \cite{Maan,M87}. This system is described by two oscillators with the frequencies  $\omega_\xi(\theta)$ and $\omega_\eta(\theta)$, where $\theta$ is a polar angle of the magnetic field \mbox{\boldmath$B$}. In a strong confinement limit, $\omega_c\ll\omega_0$, these frequencies reduce to $\omega_\xi(\theta)\approx\omega_c\cos\theta$ and $\omega_\eta(\theta)\approx\omega_0$, where $\omega_c=eB/mc$ is a cyclotron frequency in a normal field and $m$ is the effective mass. The dependence of the EDSR intensity on the spatial orientation of \mbox{{\boldmath $B$}($\theta,\varphi$)}, $\varphi$ being the azimuth of \mbox{\boldmath$B$}, can be found for BIA and SIA mechanisms by taking matrix elements of $z$ between the up- and down-spin states. For a [0,0,1] quantum well and the oscillator ground state, $n_\xi=n_\eta=0$, the result reads \cite{RE}:
\begin{eqnarray}
I_D(\theta,\varphi)&=&\left({{\alpha_D}\over{\hbar\omega_0}}\right)^2
{{\omega_c^2\omega_s^2\sin^2\theta}\over{\omega_0^2(\omega_c^2\cos^2\theta-\omega_s^2)^2}}\nonumber\\
&\times&[(\omega_c-\omega_s)^2\cos^2\theta\sin^22\phi
+(\omega_c\cos^2\theta-\omega_s)^2\cos^22\phi],\nonumber\\
I_R(\theta,\varphi)&=&\left({{\alpha_R}\over{\hbar\omega_0}}\right)^2
{{\omega_c^2\omega_s^2(\omega_c-\omega_s)^2}\over{\omega_0^2(\omega_c^2\cos^2\theta-\omega_s^2)^2}}
\sin^2\theta\cos^2\theta.
\label{eq4}
\end{eqnarray}

There are similarities between the intensities found for both mechanisms. First, the confinement frequency $\omega_0$ appearing in the denominators emphasizes that the gate controlled EDSR develops due to a deviation from the strictly 2D regime and suggests that it is desirable to keep $\omega_c\sim\omega_0$. Second, the spin frequency $\omega_s$ in the numerators suggests that EDSR should be stronger for the compounds with large $g$-factors typical of narrow gap semiconductors. Third, $\sin\theta$ in the numerators underscores that the field \mbox{\boldmath$B$} should be tilted to mix the in-plane and perpendicular-to-plane motions. Fourth, intensities reach their maxima near the resonance between the frequency of the orbital motion $\omega_\xi(\theta)\approx\omega_c\cos\theta$ and the spin-flip frequency $\omega_s$. The singularity is cut-off when the level anticrossing due to the SO interaction is taken into account but mostly by the level widths.

There are also remarkable differences between these two angular dependencies making them an unique tool for identifying the contributions to the EDSR intensity coming from various SO coupling mechanisms. First, $I_D(\theta,\varphi)$ shows a strong $\varphi$-dependence while $I_R(\theta,\varphi)$ is axially symmetric. Second, for an in-plane field \mbox{\boldmath$B$}, when $\theta=\pi/2$, the intensity $I_R(\pi/2,\varphi)=0$ while $I_D(\pi/2,\varphi)$ remains finite and shows a $\cos^22\varphi$ angular dependence.

When $\alpha_D$ and $\alpha_R$ are of a comparable magnitude, the Dresselhaus and Rashba contributions to the spin transition amplitude interfere similarly to the interference of EDSR and EPR in bulk InSb \cite{Dob83}. As a result, the four-fold symmetry of $I(\theta,\varphi)$ should be broken, just similarly to the symmetry of the energy \cite{RS91,And92} and Raman \cite{J95} spectra. Only a two-fold symmetry axis survives.

To compare the intensities of EDSR and EPR, it is convenient to introduce a characteristic length of EDSR $l_\alpha=\alpha/\hbar\omega_0$. For $\alpha\approx10^{-9}$ eV cm and $\hbar\omega_0\approx 20$ meV we find $l_\alpha\approx 5\times10^{-8}$ cm. It is three orders of magnitude larger than the Compton length $\lambda_C=\hbar/m_0c=4\times 10^{-11}$ cm that is a characteristic length for EPR. Therefore, even with the small factors like $\omega_c/\omega_0$ and $\omega_s/\omega_0$ entering into Eq.~(\ref{eq4}) and suppressing EDSR, there is a lot of space left for $l_\alpha$ to be larger than $\lambda_C$, and electrical manipulation of electron spins is preferable to magnetic one not only because it allows access to electron spins at a nanometer scale but also because a larger coupling constant can be achieved. It is also worth to emphasize that using strong magnetic fields is advantageous not only because they increase the EDSR intensity but mostly because they impose a single precession frequency on the electron ensemble and therefore suppress the D'yakonov-Perel' spin relaxation mechanism \cite{DP72}.  

\section{Electrical ringing of electron spins}
\label{ring}

The same interaction of electron spins with electric fields that allows electrical manipulating electron spins also allows noninvasive electrical detection of spin dynamics in quantum wells and quantum dots \cite{LR03}. E.g., when a dot is in a mixed quantum state, electron spin oscillates with a Zeeman frequency $\omega_s=g\mu_BB/\hbar$ between the spin-up and spin-down stationary states and produces an electric dipole and electric multipoles (side by side with a magnetic dipole) oscillating at the same frequency. For a strongly elongated (nearly 1D) quantum dot with a confinement length $L$ and a confinement frequency $\omega_0$ in the longitudinal direction, the magnitude of the electric dipole is about
\begin{equation}
 P(t)\sim {{g\mu_BB}\over{\hbar\omega_0}}~{{\alpha/L}\over{\hbar\omega_0}}~eL\cos\omega_st~.
\label{eq5}
\end{equation}
A numerical coefficient in Eq.~(\ref{eq5}) depends on the type of confinement (parabolic, rectangular), and $P(t)$ increases resonantly when $\omega_s$ approaches eigenfrequencies of electric-dipole allowed transitions. Interaction of the electric field of a precessing spin with metallic leads results in a new spin relaxation mechanism.

\section{Electric spin injection: role of resistive contacts}
\label{diffusive}

Spin injection from a ferromagnetic source (F) into a normal (non-magnetic) conductor (N) is of critical importance for spintronic devices. After efficient metal-to-metal injection has been achieved by  Johnson and Silsbee \cite{JS85}, it has been proved \cite{Fil00} that metal-to-semiconductor injection was surprisingly low. In the framework of a diffusive theory by van Son {\it et al.} \cite{vS87}, Schmidt {\it et al.} \cite{Sch00} have clarified the physical origin of the problem in terms of the conductivity mismatch. The spin injection coefficient $\gamma$ is usually defined as a ratio $\gamma=(j_\uparrow-j_\downarrow)/J$ of the spin-polarized current to the total current $J=j_\uparrow+j_\downarrow$. In convenient notations, the expression for $\gamma$ for a F-N-junction reads  \cite{HZ97,R00}
\begin{equation}
\gamma=[r_c(\Delta\Sigma/\Sigma)+r_F(\Delta\sigma/\sigma_F)]/(r_F+r_c+r_N)~. 
\label{eq6}
\end{equation}
Here $\sigma_F=\sigma_\uparrow+\sigma_\downarrow$ is the conductivity of the ferromagnet, $\Delta\sigma=\sigma_\uparrow-\sigma_\downarrow$ is a difference in the bulk conductivities of the up- and down-spins, and $r_F=\sigma_F L_F/4\sigma_\uparrow\sigma_\downarrow$ plays a role of the effective resistance of the F region, $L_F$ being the spin diffusion length in this region. The resistance $r_N=L_N/\sigma_N$ plays a similar role for the N region. The F-N-contact is described by the conductivities $\Sigma_\uparrow$ and $\Sigma_\downarrow$ for the up- and down-spins, respectively, $\Sigma=\Sigma_\uparrow+\Sigma_\downarrow$, $\Delta\Sigma=\Sigma_\uparrow-\Sigma_\downarrow$, and $r_c=\Sigma/4\Sigma_\uparrow\Sigma_\downarrow$ is the effective contact resistance.

For perfect nonresistive contacts $r_c=0$, and the magnitude of $\gamma$ is determined solely by the competition between $r_N$ and $r_F$. Because $r_N\gg r_F$ for a metal-semiconductor contact, $\gamma$ is small. This is the conductivity mismatch concept, and with nonresistive contacts only two solutions of the problem are possible: using semimagnetic semiconductors ($r_N\sim r_F$) or half-metals ($r_F=\infty$) as spin emitters. The other option is using usual ferromagnets like Fe or Permalloy and resistive contacts with $r_c>r_N$ \cite{R00}. Under these conditions $\gamma\approx\Delta\Sigma/\Sigma$, hence, high spin selectivity of the contact, i.e., large difference in the contact conductivities for up- and down-spins, is the central problem. It was solved successfully during the last two years by a number of independent experimental groups \cite{P02}.

The general conclusion is as follows. It is the element of the junction with the largest effective resistance that controls spin injection across it, and the spin injection coefficient $\gamma$ of the junction is nearly equal to the spin selectivity of this element. The underlying physics is very simple. In the regions with lower effective resistances nonequilibrium spins are accumulated at a spatial scale of about $L_F$ (or $L_N$) from the contact, and their diffusive currents fine-adjust $\gamma$ to the level imposed by the high-resistance element.

Diffusive theory of spin injection across a F-N-F-junction becomes cumbersome when written for resistive contacts. However, it can be highly simplified in the framework of the $\gamma$-technique that operates with spin injection coefficients across different contacts as the basic variables \cite{R02}.

\section{Spin injection into a ballistic region}

It is seen from Eq.~(\ref{eq6}) that it is the high effective resistance $r_N$ of the N region that suppresses spin injection. For a F-N-F-junction, the effective resistance of this region $r^*_N$ depends on the N region thickness $d$, and in the framework of a diffusive theory $r^*_N(d)\approx d/\sigma_N$ when $d$ is small enough. The problem is whether by reducing $d$ [hence, also $r^*_N(d)$] one can finally achieve high $\gamma$ without using resistive contacts?

It follows from the Sharvin resistance concept and from the Landauer-B\"{u}ttiker theory that when $d$ becomes so small that the transport in the N region is ballistic, the contribution of this region into the total resistance remains at the level of the resistance quantum $h/e^2$ per channel. The potential produced by the Sharvin resistance drops in the diffusive regions at the spatial scale of ${\it l}_F$, a mean free pass in the F regions. However, these results say nothing about $\gamma$ because the spatial scales responsible for the spin injection and the Sharvin resistance [$L_F(L_N)$ and ${\it l}_F$, respectively] are very different.

Boltzmann theory of spin injection into a ballistic N region \cite{KR03} shows that $\gamma$ is controlled by the same Sharvin resistance per unit cross-section, $r_N^*=(h/e^2)(\pi^2/S_N)$, as the ballistic region contribution to the total resistance of a F-N-F-junction. Here $S_N=\pi k_N^2$ is the cross-section of the Fermi surface and $k_N$ is the Fermi momentum in the N region. Spin injection is described by an equation similar to Eq.~(\ref{eq6}) with the resistance $r_N$ substituted by $r_N^*$. Therefore, resistive spin-selective contacts are still needed, and the condition for efficient spin injection reads as $r_c>r_N^*$ provided spin selectivity of the contact is high enough. Because $d/\sigma_N\sim h/e^2k_N^2$ when $d\sim {\it l}_N$, equations of the ballistic and diffusive theories match smoothly at $d\sim {\it l}_N$, ${\it l}_N$ being a mean free path in the N region.

These conclusions have important implications for the metal-semiconductor contacts with high single-electron transparencies that are discussed currently in the literature and are of high intrinsic interest \cite{K01,MPD00}. However, the above results suggest that a ``perfect" contact with a nearly 100\% single-electron transparency cannot solve the spin-injection problem because the suppression of spin injection into a ballistic semiconductor comes from the disparity in the sizes of the Fermi surfaces of the ferromagnetic metal and the semiconductor (a ``concentration mismatch") resulting in a strong perturbation of the electron distribution function inside the ferromagnet.

\section{Coulomb nonlocality in low-dimensional conductors}

A local Ohmic relation between the electric field \mbox{\boldmath$E$(\boldmath$r$)} and current density \mbox{\boldmath$j$(\boldmath$r$)} in regular conductors is ensured by the exponential 3D electrical screening of the electric fields proportional to the currents. In quantum conductors with a strong space quantization of electronic states the exponential screening breaks down \cite{AFS82}. The consequences of this change in the screening pattern can be comprehended from the following simple arguments. An inhomogeneous electric field developing around an inhomogeneity of a quantum conductor, according to a Poisson equation, implies accumulation of a space charge proportional to the current. Because the screening in a quantum conductor is only of the power-law type, the electric field of this space charge penetrates into the remote sections of the same conductor, and even into the different quantum conductors, and influences the currents flowing through them. This long range interaction results in a Coulomb nonlocality in the conductivity, current-dependent mechanical forces between different quantum conductors, etc.

Another phenomenon related to these long-range current-induced electric fields is their large electrostatic energy and, hence, the giant capacitances that are cut-off only by the screening of the fields by  classical 3D conductors surrounding a quantum conductor \cite{KR02}. E.g., a resistive contact inside a 2D quantum conductor acquires capacitance  $C\approx(\varepsilon/2\pi^2)\ln(L/{\it l}_{2D})$ per unit length. Here ${\it l}_{2D}$ is a 2D screening length, $\varepsilon$ is a dielectric constant, and $L$ is a distance of the contact from 3D conductors. Giant capacitances are a classical phenomenon accompanying quantum transport. They should affect performance of quantum circuits at high frequencies, in particular, of spin injecting contacts.

The above results are based on explicit solving the Coulomb problem for the dc and ac transport across quantum conductors with simple geometries and clarify some new aspects of the macroscopic theory of mesoscopic networks \cite{But93}.

Support from DARPA/SPINS by the Office of Naval Research Grant N000140010819 is gratefully acknowledged.

\end{document}